\documentclass[runningheads]{llncs}
\usepackage{graphicx}
\usepackage{multirow}
\usepackage{amsfonts}
\usepackage{amsmath}
\DeclareMathOperator{\Tr}{Tr}
\usepackage{siunitx}
\usepackage{hyperref}

\begin{document}
\title{Conditional Generative Data Augmentation for Clinical Audio Datasets}
%
%

\author{Matthias Seibold \inst{1,2} \and
Armando Hoch \inst{3} \and
Mazda Farshad \inst{3} \and
*Nassir Navab \inst{1}\and
*Philipp Fürnstahl \inst{2,3}
}


\institute{Computer Aided Medical Procedures (CAMP), 
Technical University of Munich, Munich, DE-85748, Germany \and
Research in Orthopedic Computer Science (ROCS), 
University Hospital Balgrist, University of Zurich, Zurich, CH-8008, Switzerland \and
Balgrist University Hospital, Zurich, CH-8008, Switzerland\\
*equally contributing last authors}

\authorrunning{Seibold et al.}

%
%

\maketitle              
\begin{abstract}
In this work, we propose a novel data augmentation method for clinical audio datasets based on a conditional Wasserstein Generative Adversarial Network with Gradient Penalty (cWGAN-GP), operating on log-mel spectrograms. To validate our method, we created a clinical audio dataset which was recorded in a real-world operating room during Total Hip Arthroplasty (THA) procedures and contains typical sounds which resemble the different phases of the intervention. We demonstrate the capability of the proposed method to generate realistic class-conditioned samples from the dataset distribution and show that training with the generated augmented samples outperforms classical audio augmentation methods in terms of classification performance. The performance was evaluated using a ResNet-18 classifier which shows a mean Macro F1-score improvement of $1.70\%$ in a 5-fold cross validation experiment using the proposed augmentation method. Because clinical data is often expensive to acquire, the development of realistic and high-quality data augmentation methods is crucial to improve the robustness and generalization capabilities of learning-based algorithms which is especially important for safety-critical medical applications. Therefore, the proposed data augmentation method is an important step towards improving the data bottleneck for clinical audio-based machine learning systems.

\end{abstract}
\section{Introduction}

Acoustic signals are easy and low-cost to acquire, can be captured using air-borne or contact microphones and show great potentials in medical applications for interventional guidance and support systems. Successful applications are intra-operative tissue characterization during needle insertion \cite{illanes2018novel} and tissue coagulation \cite{ostler2020acoustic}, the identification of the insertion endpoint in THA procedures \cite{goossens2020acoustic,seibold2021femoralstem}, error prevention in surgical drilling during orthopedic procedures \cite{seibold2021realtime} or guidance in orthopedic arthroscopy procedures \cite{suehn2020acoustic}.

Furthermore, acoustic signals have successfully been employed for diagnostic medical applications. Exemplary applications include the assessment of cartilage degeneration by measuring structure borne noise in the human knee during movement \cite{kim2009vag}, the development of a prototype for the detection of implant loosening through an acoustic sensor system \cite{ewald2011acoustic}, a system for monitoring the acoustic emissions of THA implants \cite{rodgers2014emission}, or the automated analysis of lung sounds captured with a digital stethoscope which allows non-specialists to screen for pulmonary fibrosis \cite{marshall2007chest}.

Through recent advances in machine learning research, learning-based methods have replaced and outperformed classical acoustic signal processing-based approaches, as well as classical handcrafted feature-based learning approaches for many acoustic audio signal processing tasks \cite{purwins2019deep}. However, state-of-the-art deep learning methods require large amounts of training data to achieve superior performance and generalize well to unseen data, which are often difficult or infeasible to acquire in a clinical setting. To tackle this issue, the usage of augmentation techniques is a standard approach to increase the diversity and size of training datasets. Hereby, new samples can be synthesized by applying transformations to the existing data, e.g. rotation and cropping for images, replacing words with synonyms for text, and applying noise, pitch shifting, and time stretching to audio samples \cite{wei2020audioaugmentation}. Even though these data augmentation methods improve the performance of target applications, they do not necessarily generate realistic samples which is especially crucial in the medical domain where reliability is a key factor. One solution to this problem is for example to exploit the underlying physics for augmentation, e.g. for ultrasound image augmentation \cite{tirindelli2021ultrasound} which is, however, not applicable for clinical audio data. In the presented work, we will focus on realistic data augmentation of audio datasets for medical applications. 

Recently, deep generative models, a family of deep learning models, which are able to synthesize realistic samples from a learned distribution, have been applied for data augmentation of various data modalities outside of the medical domain. For the augmentation of audio data, different generative approaches have been introduced, of which related work to the proposed method is described in the following section. Hu et al. utilized a GAN to synthesize samples of logarithmic Mel-filter bank coefficients (FBANK) from a learned distribution of a speech dataset and subsequently generated soft labels using a pretrained classifier \cite{hu2018gan}. Madhu et al. trained separate GANs on mel-spectrograms for each class of a dataset to generate augmentation data \cite{madhu2019augment}. Chatziagapi et al. used the Balancing GAN (BAGAN) framework \cite{Mariani2018BAGANDA} to augment an imbalanced speech dataset \cite{Chatziagapi2019}. A conditional GAN was employed for data augmentation of speech using FBANK features by Sheng et al. \cite{sheng2018augmentation} and for respiratory audio signals based on raw waveform augmentation by Jayalakshmy et al. \cite{jayalakshmy2021conditionalgan}. 

In this work, we introduce a novel augmentation technique for audio data based on a conditional Wasserstein GAN model with Gradient Penalty (cWGAN-GP) which produces higher-quality samples and is easier to train than standard GANs \cite{gulrajani2017improved}. The proposed model operates on log-mel spectrograms which have been shown to outperform other feature representations and achieves state-of-the-art performance in audio classification tasks \cite{purwins2019deep}. 
The proposed model is able to generate realistic and high-quality log-mel spectrograms from the learned dataset distribution. We show that our model can be used for two augmentation strategies, doubling the number of samples and balancing the dataset. While classical audio augmentation techniques might improve the performance of the classifier, they do not generate samples that can be captured in a real environment and might therefore be inconsistent with the real variability of captured real-world acoustic signals. In contrast, the proposed model is able to generate realistic samples from the learned distribution of the original data.

To evaluate the proposed framework on realistic clinical data, we introduce a novel audio dataset containing sounds of surgical actions recorded from five real THA procedures which resemble the different phases of the intervention. We thoroughly evaluate the proposed method on the proposed dataset in terms of classification performance improvement of a ResNet-18 classifier with and without data augmentation using 5-fold cross validation and compare the results with classical audio augmentation techniques.

\section{Materials and Method}

\subsection{Novel Surgical Audio Dataset}

\begin{figure}[ht]
    \centering
    \includegraphics[width=\textwidth]{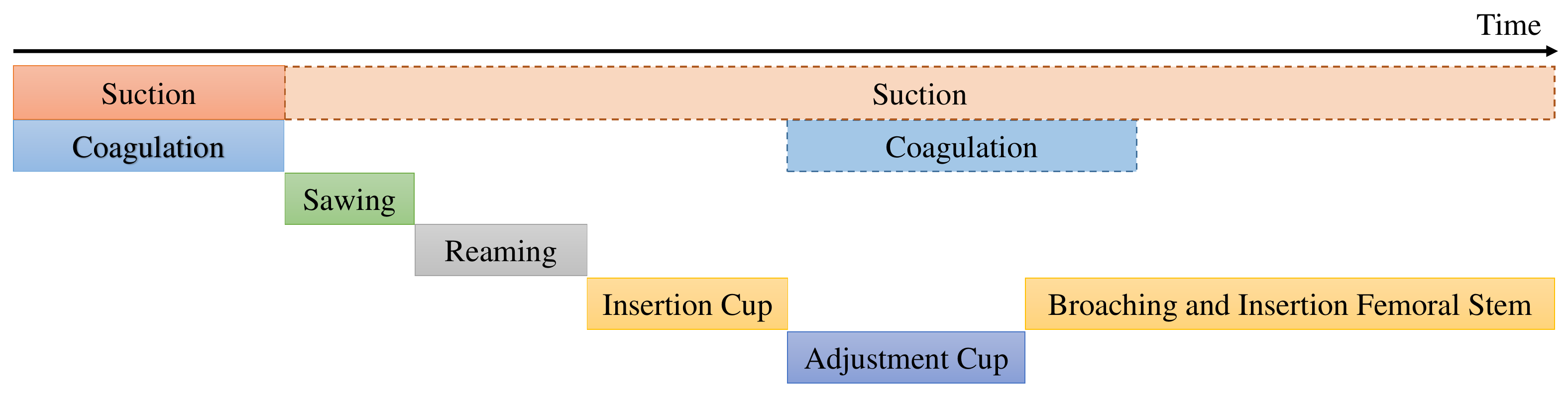}
    \caption{The classes of the novel clinical dataset resemble the phases of a THA procedure. Occurrences with drawn through lines indicate intensive usage of the respective surgical action, dashed lines correspond to sporadic usage.}
    \label{fig:tha}
\end{figure}

Figure \ref{fig:tha} illustrates the occurrence of the six classes $C:=$ \{Suction, Coagulation, Sawing, Reaming, Insertion, Adjustment\} present in the dataset over the course of a THA procedure. Please note that "Insertion Cup" and "Broaching and Insertion Femoral Stem" were joined into a single class ("Insertion") because of the similar acoustic signature generated by hammering onto the metal structure of the insertion tools for the acetabular cup, femoral broach and femoral stem implant, respectively. The "Adjustment" class also contains hammering signals that are, however, performed with a screwdriver-like tool which is used to adjust the orientation of the acetabular cup and generates a slightly different sound. During opening the access to the area of operation in the beginning of the procedure, suction and coagulation is employed intensively, whereas in the rest of the procedure both surgical actions are performed sporadically and on demand (indicated through dashed outlines in Figure \ref{fig:tha}). All samples were manually cut from recordings of five THA interventions conducted at our university hospital for which we captured audio with a framerate of \SI{44.1}{\kilo\Hz} using a air-borne shotgun microphone (R{\o}de NTG2) pointed towards the area of operation and video captured from the OR light camera (Trumpf TruVidia). The captured video was used to facilitate the labelling process. We labelled the dataset in a way that audio samples do not contain overlapping classes and no staff conversations. An ethical approval has been obtained prior to recording the data in the operating room. The resulting dataset contains 568 recordings with a length of \SIrange{1}{31}{\s} and the following distribution: $n_{raw, Adjustment} = 68$, $n_{raw, Coagulation} = 117$, $n_{raw, Insertion} = 76$, $n_{raw, Reaming} = 64$, $n_{raw, Sawing} = 21$, and $n_{raw, Suction} = 222$. The dataset can be accessed under \href{https://rocs.balgrist.ch/en/open-access/}{https://rocs.balgrist.ch/en/open-access/}.

\subsection{Data Preprocessing and Baseline Augmentations}
\label{subsec:data}

Log-mel spectrograms are a two-dimensional representation of an audio signal, mapping frequency components of a signal to the ordinate and time to the abscissa. They offer a dense representation of the signal, reduce the dimensionality of the samples, and have been shown to yield superior classification performance for a wide variety of acoustic sensing tasks \cite{purwins2019deep}. We compute log-mel spectrograms of size $64\times64$ from the dataset samples by applying a sliding window technique with non-overlapping windows of length $L = 16380$ samples, a Short Time Fourier Transform (STFT) hop length of $H = 256$ samples and $n_{mels} = 64$ mel bins using the Python library \textit{librosa 0.8.1} \cite{mcfee2015librosa}. We compute a total number of 3597 individual spectrograms from the raw waveform dataset. The resulting number of spectrograms per-class is: $n_{spec, Adjustment} = 494$, $n_{spec, Coagulation} = 608$, $n_{spec, Insertion} = 967$, $n_{spec, Reaming} = 469$, $n_{spec, Sawing} = 160$, and $n_{spec, Suction} = 899$.
For the evaluation using 5-fold cross validation, we randomly split the dataset into five folds on the raw waveform level over all recordings, as the recording conditions are identical.

To compare the proposed augmentation method against classical signal processing augmentation approaches, we implemented the following augmentation strategies which are applied to the raw waveforms directly. We apply Gaussian noise with $\mu = 0$ and $\sigma = 0.01$. We apply Pitch Shifting by 3 semitones upwards. We apply time stretching with a factor of 1.5. Furthermore, we compare our method with \textit{SpecAugment}, a widely used approach for audio augmentation in Automatic Speech Recognition (ASR) tasks which applies random time-warping, frequency- and time-masking to the spectrograms directly \cite{park2019specaugment}. For a fair comparison of all augmentations, we add 100\% generated samples for each augmentation strategy, respectively. We normalize the data by computing $X_{norm, mel} = {(X_{mel} - \mu )} / {\sigma}$, where ($\mu$) is the mean and ($\sigma$) is the standard deviation of the entire dataset.

\subsection{Conditional Generative Data Augmentation Method}

\begin{figure}[ht]
    \centering
    \includegraphics[width=\textwidth]{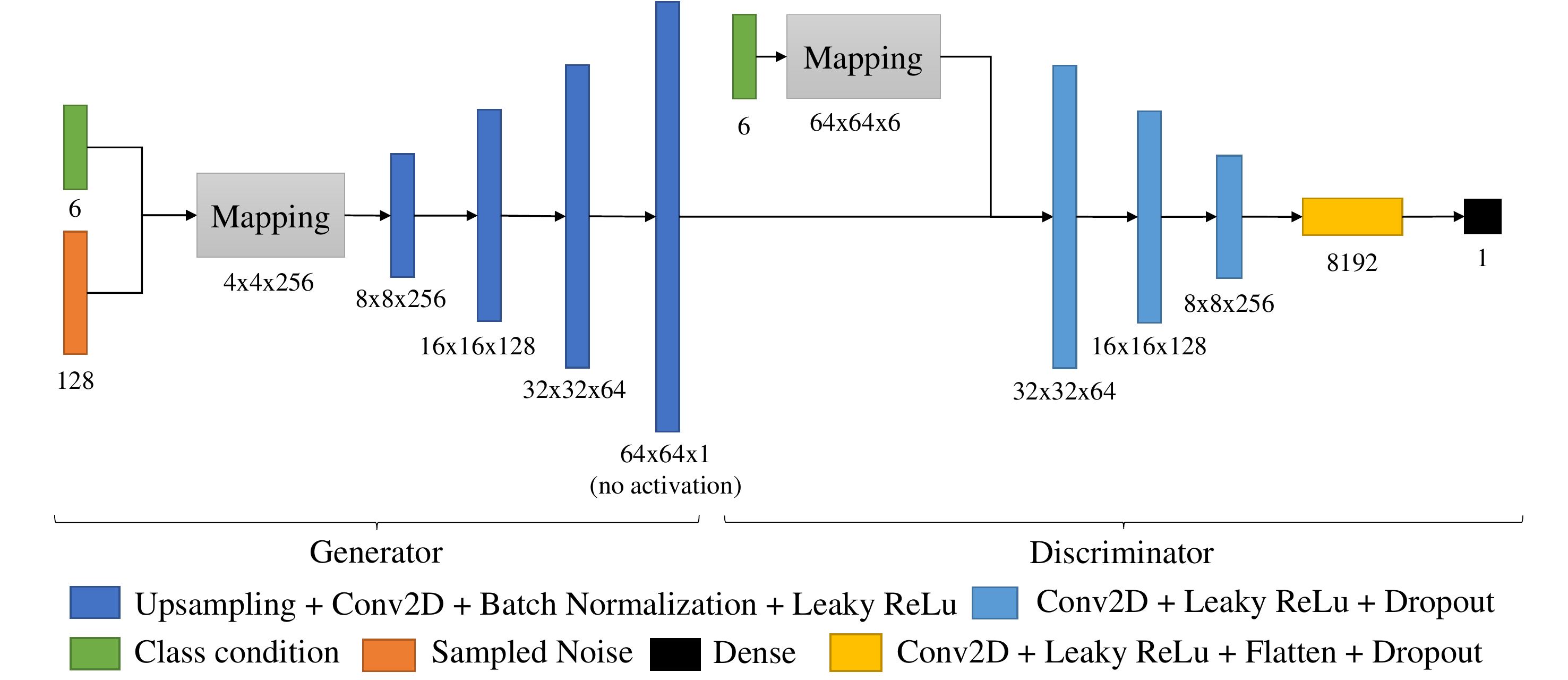}
    \caption{The architecture of the proposed model including output sizes of each layer. The input for the generator is a noise vector of size 1x128 and a class condition. The generator outputs a spectrogram which is fed to the discriminator together with the class condition. The discriminator (critic) outputs a scalar realness score.}
    \label{fig:cwgangp}
\end{figure}

The architectural details of the proposed GAN are illustrated in Figure \ref{fig:cwgangp}. To stabilize the training process and improve the generated sample quality, we apply the Wasserstein loss with Gradient Penalty (GP) as introduced by Gulrajani et al. \cite{gulrajani2017improved} which enforces a constraint such that the gradients of the discriminator's (critic) output w.r.t the inputs have unit norm. This approach greatly improves the stability of the training and compensates for problems such as mode collapse. We define the critic's loss function as:

\begin{equation}
    L_C = \mathbb{E}_{\tilde{x}\sim\mathbb{P}_g}[D(\tilde{x},y)] - \mathbb{E}_{x\sim\mathbb{P}_r}[D(x,y)] + \lambda \; \mathbb{E}_{\hat{x}\sim\mathbb{P}_{\hat{x}}}[\|\Delta_{\hat{x}}D(\hat{x},y)\|_2 - 1)^2]
\end{equation}

where $\mathbb{P}_r$ is the real distribution, $\mathbb{P}_g$ is the generated distribution, and $\mathbb{P}_{\hat{x}}$ is the interpolated distribution. The interpolated samples $\hat{x}$ are uniformly sampled along a straight line between real $x$ and generated $\tilde{x}$ samples by computing:

\begin{equation}
    \hat{x} = \epsilon x + (1-\epsilon) \tilde{x}
\end{equation}

We use the recommended GP weight of $\lambda = 10$, a batch size of 64 and train the discriminator five times for each generator iteration. In order to choose the stopping point for training, we frequently compute the Fréchet Inception Distance (FID) \cite{heusel2017fid} which is calculated from features of a pretrained classifier by:

\begin{equation}
    FID = \|\mu_r - \mu_g\|^2 + \Tr(C_r + C_g - 2*\sqrt{C_r * C_g})
\end{equation}

as a measure of the quality for the generated samples and stop the training at epoch 580. Hereby, $\mu_r$ and $\mu_g$ represent the feature-wise mean of the real and generated spectrograms, $C_r$ and $C_g$ the respective covariance matrices. Because of the structural differences of images and spectrograms, we cannot use an Inception v3 network pretrained on ImageNet to compute the FID. Therefore, we employ a ResNet-18 \cite{he2016resnet} model pretrained on the proposed dataset, extract the features from the last convolutional layer, and use these features for FID calculation. The proposed model is implemented with \textit{TensorFlow/Keras 2.6} and trained using the Adam optimizer ($LR = 1e-4$, $\beta_1 = 0.5$, $\beta_2 = 0.9$) in $\sim$6 hours on a NVidia RTX 2080 SUPER GPU.

A nonlinear activation function is omitted in the last convolutional layer of the generator because the spectrogram samples are not normalized in the range $[0, 1]$. The mapping layer of the generator employs a dense layer, whereas in the discriminator (critic) we use repeat and reshaping operations for remapping. The generator and discriminator have a total number of 1,526,084 and 4,321,153 parameters, respectively. The implementation, pretrained models, and dataset can be accessed under: \href{https://rocs.balgrist.ch/en/open-access/}{https://rocs.balgrist.ch/en/open-access/}.

\subsection{Classification Model}

To evaluate the  augmentation performance of our model against  classical audio augmentation techniques, we analyze the effect of augmentation on the classification performance in a 5-fold cross validation experiment using a ResNet-18 \cite{he2016resnet} classifier, an architecture which has been successfully employed for clinical audio classification tasks \cite{seibold2021femoralstem,seibold2021realtime}. We train the classifier from scratch for 20 epochs using categorical crossentropy loss, the Adam optimizer ($LR = 1e-4$, $\beta_1 = 0.9$, $\beta_2 = 0.999$) and a batch size of 32.

\section{Results and Evaluation}

\begin{figure}[ht]
    \centering
    \includegraphics[width=\textwidth]{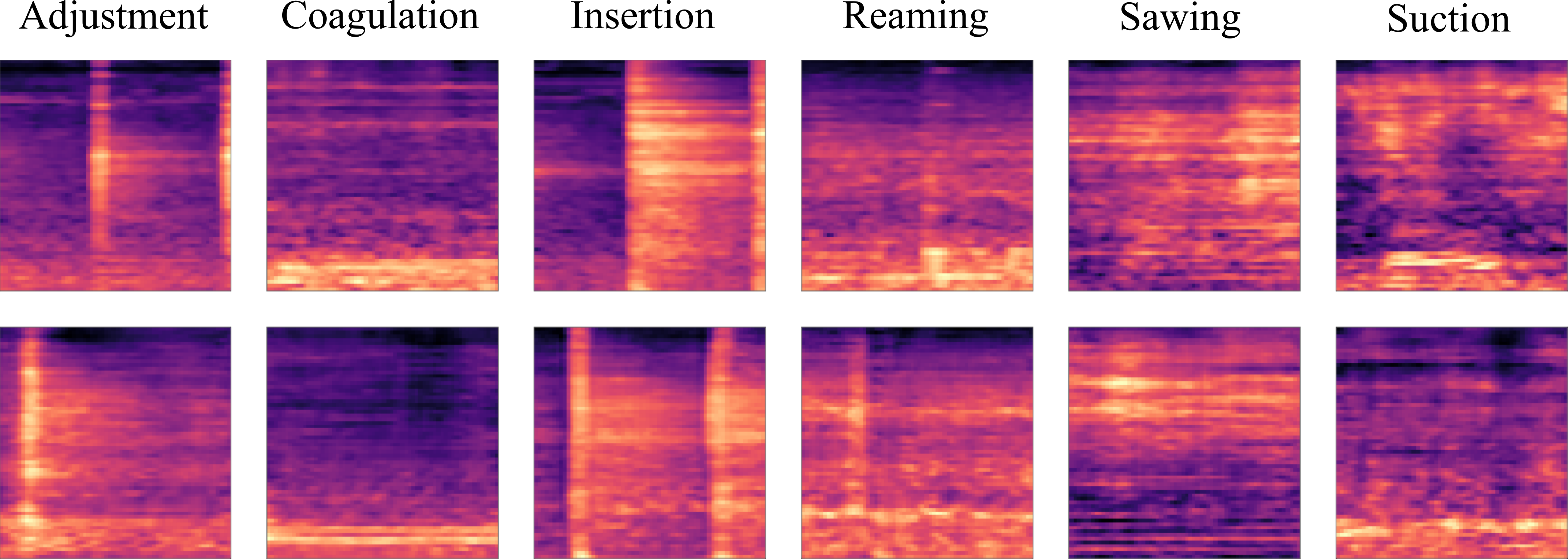}
    \caption{The top row shows log-mel spectrograms of random samples for each class present in the acquired dataset, the bottom row shows log-mel spectrograms generated by the proposed model for each class, respectively.}
    \label{fig:generated}
\end{figure}

In Figure \ref{fig:generated}, we show a comparison of randomly chosen ground truth and randomly generated samples. The visual quality of the generated samples is comparable to the original data and the model seems to be able to generate samples conditioned on the queried class. By further visual inspection it can be observed that the synthesized samples contain the characteristics of the original dataset, e.g. the hammer strokes are clearly visible for the classes "Adjustment" and "Insertion". 

The quantitative evaluation of the classification performance using a ResNet-18 classifier is given in Table \ref{tab:results}. We report the mean Macro F1-Score in the format $mean \pm std$. . We compare training without augmentations and classical audio augmentation techniques (adding noise, pitch shifting, time stretching, and SpecAugment \cite{park2019specaugment}) with the proposed method. The cWGAN-GP-based augmentations outperform all classical augmentation strategies when doubling the samples ($+1.70 \%$) and show similar performance ($+1.07 \%$) as the best performing classical augmentation strategy (Time Stretch) when balancing the dataset. 

\begin{table}
    \centering
    \begin{tabular}{|p{0.3\textwidth}|p{0.25\textwidth}|p{0.2\textwidth}|}
    \hline
        \textbf{Augmentation \quad\quad\quad Technique}   &   \textbf{Mean Macro\newline F1-Score}    &   \textbf{Relative \quad\quad\quad Improvement}\\
        \hline\hline
        No Augmentation     &   $93.90 \pm 2.48 \%$  &              \\
        Add Noise           &   $92.87 \pm 0.99 \%$  &   $-1.03 \%$ \\
        Pitch Shift         &   $94.73 \pm 1.28 \%$  &   $+0.83 \%$ \\
        Time Stretch        &   $95.00 \pm 1.49 \%$  &   $+1.10 \%$ \\
        SpecAugment \cite{park2019specaugment}         &   $94.23 \pm 1.14 \%$  &   $+0.33 \%$\\
        cWGAN-GP (balanced) &   $94.97 \pm 1.71 \%$  &   $+1.07 \%$ \\
        cWGAN-GP (doubled)  &   $\mathbf{95.60 \pm 1.26 \%}$  &   $\mathbf{+1.70 \%}$ \\
        \hline
    \end{tabular}
    \caption{Results of the proposed model in comparison to classical audio augmentation techniques.}
    \label{tab:results}
\end{table}

\section{Discussion}

The proposed augmentation method is an important step towards improving the data limitations by generating synthetic in-distribution augmentation data for clinical applications for which it is often expensive or even impossible to gather large amounts of training data. 
We showed that our augmentation strategy outperforms classical signal processing approaches and has the capability to balance imbalanced datasets to a certain extent. To balance imbalanced datasets, any arbitrary number of samples can easily be generated for each class with the proposed approach which is not possible using classical signal processing techniques in the same way. However, for the given dataset and configuration, doubling the number of samples using the proposed augmentation method leads to the best final classification results. Furthermore, we show that the proposed method outperforms SpecAugment \cite{park2019specaugment}, an established audio augmentation method which applies time-warping, as well as frequency and time masking to the spectrogram data directly. 

In future work, we want to benchmark the proposed framework with other generative augmentation models and model architectures, investigate the performance of the proposed approach on more (balanced and imbalanced) datasets and further optimize our model towards improved classification performance. Furthermore, it should be investigated how combinations of augmentation techniques influence the resulting classification performance and if it is possible to maximize the impact of augmentations through an optimized combination scheme.

The improved performance achieved by the proposed augmentation method comes at the cost of increased demands on computational power and resources. While signal processing augmentations are computed in the range of seconds to minutes, our model requires an additional training step which takes $\sim$6 hours for the presented dataset and increases with larger datasets. 

Because we created the proposed clinical dataset in a way that it resembles the phases and surgical actions executed during a real THA procedure, potential future clinical applications are the prediction of surgical actions from captured audio signals in the operating room which could be used for workflow recognition and surgical phase detection. Therefore, we consider the proposed dataset as an important step towards automated audio-based clinical workflow detection systems, a topic which has only been studied rudimentally so far \cite{suzuki2010workflow,weede2012workflow}. The proposed approach is designed to work with spectrogram based audio, which can be transformed back to the signal domain, e.g. using the Griffin-Lim algorithm \cite{griffin1984algo} or more recently introduced learning-based transformation approaches, e.g. the work by Takamichi et al. \cite{takamichi2020wavreconstruction}. We reconstructed waveforms from a few generated spectrograms using the Griffin-Lim algorithm and could, despite artifacts being present, recognize acoustic similarities to the original samples for each class, respectively. In future work, the proposed augmentation method could furthermore be transferred to other medical and non-medical grid-like data domains.

\section{Conclusion}

In the presented work, we introduce a novel data augmentation method for medical audio data and evaluate it on a clinical dataset which was recorded in real-world Total Hip Arthroplasty (THA) surgeries. The proposed dataset contains sound samples of six surgical actions which resemble the different phases of a THA intervention. We show in quantitative evaluations that the proposed method outperforms classical signal and spectrogram processing-based augmentation techniques in terms of Mean Macro F1-Score, evaluated using a ResNet-18 classifier in a 5-fold cross validation experiment. By generating high-quality in-distribution samples for data augmentation, our method has the potential to improve the data bottleneck for acoustic learning-based medical support systems.

\section*{Acknowledgment}

This work is part of the SURGENT project under the umbrella of Hochschulmedizin Zürich.

%
%
%

\bibliographystyle{splncs04}
\bibliography{bibliography}

\begin{thebibliography}{10}
\providecommand{\url}[1]{\texttt{#1}}
\providecommand{\urlprefix}{URL }
\providecommand{\doi}[1]{https://doi.org/#1}

\bibitem{Chatziagapi2019}
Chatziagapi, A., Paraskevopoulos, G., Sgouropoulos, D., Pantazopoulos, G.,
  Nikandrou, M., Giannakopoulos, T., Katsamanis, A., Potamianos, A., Narayanan,
  S.: {Data Augmentation Using GANs for Speech Emotion Recognition}. In: Proc.
  Interspeech 2019. pp. 171--175 (2019)

\bibitem{ewald2011acoustic}
Ewald, H., Timm, U., Ruther, C., Mittelmeier, W., Bader, R., Kluess, D.:
  Acoustic sensor system for loosening detection of hip implants. In: 2011
  Fifth International Conference on Sensing Technology. pp. 494--497 (2011)

\bibitem{goossens2020acoustic}
Goossens, Q., Pastrav, L., Roosen, J., Mulier, M., Desmet, W., Vander~Sloten,
  J., Denis, K.: Acoustic analysis to monitor implant seating and early detect
  fractures in cementless tha: An in vivo study. Journal of Orthopedic Research
   (2020)

\bibitem{griffin1984algo}
Griffin, D., Lim, J.: Signal estimation from modified short-time fourier
  transform. IEEE Transactions on Acoustics, Speech, and Signal Processing
  \textbf{32}(2),  236--243 (1984)

\bibitem{gulrajani2017improved}
Gulrajani, I., Ahmed, F., Arjovsky, M., Dumoulin, V., Courville, A.: Improved
  training of wasserstein gans. In: Proceedings of the 31st International
  Conference on Neural Information Processing Systems. pp. 5769--–5779 (2017)

\bibitem{he2016resnet}
He, K., Zhang, X., Ren, S., Sun, J.: Deep residual learning for image
  recognition. In: 2016 IEEE Conference on Computer Vision and Pattern
  Recognition (CVPR). pp. 770--778 (2016)

\bibitem{heusel2017fid}
Heusel, M., Ramsauer, H., Unterthiner, T., Nessler, B., Hochreiter, S.: Gans
  trained by a two time-scale update rule converge to a local nash equilibrium.
  In: Proceedings of the 31st International Conference on Neural Information
  Processing Systems. p. 6629–6640 (2017)

\bibitem{hu2018gan}
Hu, H., Tan, T., Qian, Y.: Generative adversarial networks based data
  augmentation for noise robust speech recognition. In: 2018 IEEE International
  Conference on Acoustics, Speech and Signal Processing (ICASSP). pp.
  5044--5048 (2018)

\bibitem{illanes2018novel}
Illanes, A., Boese, A., Maldonado, I., Pashazadeh, A., Schaufler, A., Navab,
  N., Friebe, M.: Novel clinical device tracking and tissue event
  characterization using proximally placed audio signal acquisition and
  processing. Scientific Reports  \textbf{8} (2018)

\bibitem{jayalakshmy2021conditionalgan}
Jayalakshmy, S., Sudha, G.F.: Conditional gan based augmentation for predictive
  modeling of respiratory signals. Computers in Biology and Medicine
  \textbf{138},  104930 (2021)

\bibitem{kim2009vag}
Kim, K.S., Seo, J.H., Kang, J.U., Song, C.G.: An enhanced algorithm for knee
  joint sound classification using feature extraction based on time-frequency
  analysis. Computer Methods and Programs in Biomedicine  \textbf{94}(2),
  198--206 (2009)

\bibitem{madhu2019augment}
Madhu, A., Kumaraswamy, S.: Data augmentation using generative adversarial
  network for environmental sound classification. In: 2019 27th European Signal
  Processing Conference (EUSIPCO) (2019)

\bibitem{Mariani2018BAGANDA}
Mariani, G., Scheidegger, F., Istrate, R., Bekas, C., Malossi, A.C.I.: Bagan:
  Data augmentation with balancing gan. arXiv  \textbf{abs/1803.09655} (2018)

\bibitem{marshall2007chest}
Marshall, A., Boussakta, S.: Signal analysis of medical acoustic sounds with
  applications to chest medicine. Journal of the Franklin Institute
  \textbf{344}(3),  230--242 (2007)

\bibitem{mcfee2015librosa}
McFee, B., Raffel, C., Liang, D., PW~Ellis, D., McVicar, M., Battenberg, E.,
  Nieto, O.: librosa: Audio and music signal analysis in python. In: 14th
  python in science conference. pp. 18--25 (2015)

\bibitem{ostler2020acoustic}
Ostler, D., Seibold, M., Fuchtmann, J., Samm, N., Feussner, H., Wilhelm, D.,
  Navab, N.: Acoustic signal analysis of instrument–tissue interaction for
  minimally invasive interventions. International Journal of Computer Assisted
  Radiology and Surgery  (2020)

\bibitem{park2019specaugment}
Park, D.S., Chan, W., Zhang, Y., Chiu, C.C., Zoph, B., Cubuk, E.D., Le, Q.V.:
  Specaugment: A simple data augmentation method for automatic speech
  recognition. Interspeech 2019  (Sep 2019)

\bibitem{purwins2019deep}
Purwins, H., Li, B., Virtanen, T., Schlüter, J., Chang, S.y., Sainath, T.:
  Deep learning for audio signal processing. IEEE Journal on Selected Topics in
  Signal Processing  \textbf{14},  206--219 (2019)

\bibitem{rodgers2014emission}
Rodgers, G.W., Young, J.L., Fields, A.V., Shearer, R.Z., Woodfield, T.B.F.,
  Hooper, G.J., Chase, J.G.: Acoustic emission monitoring of total hip
  arthroplasty implants. IFAC Proceedings Volumes  \textbf{47}(3),  4796--4800
  (2014), 19th IFAC World Congress

\bibitem{seibold2021femoralstem}
Seibold, M., Hoch, A., Suter, D., Farshad, M., Zingg, P.O., Navab, N.,
  Fürnstahl, P.: Acoustic-based spatio-temporal learning for press-fit
  evaluation of femoral stem implants. In: International Conference on Medical
  Image Computing and Computer-Assisted Intervention. pp. 447--456 (2021)

\bibitem{seibold2021realtime}
Seibold, M., Maurer, S., Hoch, A., Zingg, P., Farshad, M., Navab, N.,
  Fürnstahl, P.: Real-time acoustic sensing and artificial intelligence for
  error prevention in orthopedic surgery. Scientific Reports  \textbf{11}
  (2021)

\bibitem{sheng2018augmentation}
Sheng, P., Yang, Z., Hu, H., Tan, T., Qian, Y.: Data augmentation using
  conditional generative adversarial networks for robust speech recognition.
  In: 2018 11th International Symposium on Chinese Spoken Language Processing
  (ISCSLP). pp. 121--125 (2018)

\bibitem{suehn2020acoustic}
Suehn, T., Pandey, A., Friebe, M., Illanes, A., Boese, A., Lohman, C.: Acoustic
  sensing of tissue-tool interactions – potential applications in
  arthroscopic surgery. Current Directions in Biomedical Engineering
  \textbf{6} (2020)

\bibitem{suzuki2010workflow}
Suzuki, T., Sakurai, Y., Yoshimitsu, K., Nambu, K., Muragaki, Y., Iseki, H.:
  Intraoperative multichannel audio-visual information recording and automatic
  surgical phase and incident detection. In: 2010 Annual International
  Conference of the IEEE Engineering in Medicine and Biology. pp. 1190--1193
  (2010)

\bibitem{takamichi2020wavreconstruction}
Takamichi, S., Saito, Y., Takamune, N., Kitamura, D., Saruwatari, H.: Phase
  reconstruction from amplitude spectrograms based on directional-statistics
  deep neural networks. Signal Processing  \textbf{169},  107368 (2020)

\bibitem{tirindelli2021ultrasound}
Tirindelli, M., Eilers, C., Simson, W., Paschali, M., Azampour, M.F., Navab,
  N.: Rethinking ultrasound augmentation: A physics-inspired approach. In:
  Medical Image Computing and Computer Assisted Intervention. pp. 690--700
  (2021)

\bibitem{weede2012workflow}
Weede, O., Dittrich, F., Wörn, H., Jensen, B., Knoll, A., Wilhelm, D.,
  Kranzfelder, M., Schneider, A., Feussner, H.: Workflow analysis and surgical
  phase recognition in minimally invasive surgery. In: 2012 IEEE International
  Conference on Robotics and Biomimetics (ROBIO). pp. 1080--1074 (2012)

\bibitem{wei2020audioaugmentation}
Wei, S., Zou, S., Liao, F., Lang, W.: A comparison on data augmentation methods
  based on deep learning for audio classification. Journal of Physics:
  Conference Series  \textbf{1453}(1),  012085 (2020)

\end{thebibliography}

\end{document}